\documentclass[seceq,supplement]{ptptex}

\usepackage{graphicx}
\usepackage{graphics}
\usepackage{epsfig}

\title{
How does the Smaller Alignment Index (SALI) distinguish order from
chaos? }

\author{
Charalampos \textsc{Skokos}$^{1,2,}$\footnote{E-mail:
hskokos@cc.uoa.gr}, Chris
\textsc{Antonopoulos}$^{1,}$\footnote{E-mail:
antonop@math.upatras.gr}, Tassos C.\
\textsc{Bountis}$^{1,}$\footnote{E-mail: bountis@math.upatras.gr}
and Michael N.\ \textsc{Vrahatis}$^{3,}$\footnote{E-mail:
vrahatis@math.upatras.gr} }

\inst{
$^1$Department of Mathematics, Division of Applied Analysis and
Center for Research and Applications of Nonlinear Systems (CRANS),
University of Patras, GR-26500 Patras, Greece\\
$^2$Research Center for Astronomy, Academy of Athens,
14 Anagnostopoulou str.,  GR-10673 Athens, Greece\\
$^3$Department of Mathematics
and University of Patras Artificial Intelligence
 Research Center (UPAIRC),
University of Patras, GR-26110 Patras, Greece
}

\abst{ The ability of the  Smaller Alignment Index (SALI) to
distinguish chaotic from ordered motion, has been demonstrated
recently in several publications.\cite{Sk01,GRACM} Basically it is
observed that in chaotic regions the SALI goes to zero very
rapidly, while it fluctuates around a nonzero value in ordered
regions. In this paper, we make a first step forward explaining
these results by studying in detail the evolution of small
deviations from regular orbits lying on the invariant tori of an
{\bf integrable} 2D Hamiltonian system. We show that, in general,
any two initial deviation vectors will eventually fall on the
``tangent space'' of the torus, pointing in different directions
due to the different dynamics of the $2$ integrals of motion,
which  means that the SALI (or the smaller angle between these
vectors) will oscillate away from zero for all time.}

\begin{document}

\maketitle

\section{Introduction}
The evaluation of the {\bf Smaller Alignment Index (SALI)} is an
efficient and simple  method to  determine the ordered or chaotic
nature of orbits in dynamical systems. The SALI was proposed in
Ref.~\citen{Sk01} and it has been successfully applied to
distinguish between ordered and chaotic motion both in symplectic
maps \cite{Sk01} as well as in Hamiltonian flows.\cite{GRACM}

In order to compute the SALI for a given orbit one has to follow
the time evolution of the orbit itself and  two deviation vectors
which initially point in two different directions. The evolution
of these vectors is given by the variational equations for a flow
and by the tangent map for a discrete--time system. At every time
step the two vectors $\overrightarrow{v_1}(t)$,
$\overrightarrow{v_2}(t)$ are normalized and the SALI is computed
 as:
\begin{equation}
SALI(t)= \min \left\{ \left\|
\frac{\overrightarrow{v_1}(t)}{\|\overrightarrow{v_1}(t)\|}+
\frac{\overrightarrow{v_2}(t)}{\|\overrightarrow{v_2}(t)\|} \right\|
,\left\| \frac{\overrightarrow{v_1}(t)}{\|\overrightarrow{v_1}(t)\|}
-\frac{\overrightarrow{v_2}(t)}{\|\overrightarrow{v_2}(t)\|}
\right\| \right\}, \label{eq:SALI}
\end{equation}
where $t$ is the continuous or the discrete  time and $\|\cdot\|$
denotes the Euclidean norm.

The properties of time evolution of the SALI clearly distinguish
between ordered and chaotic motion as follows: In the case of
Hamiltonian flows or $N$--dimensional symplectic maps with
$N\geqslant 2$ the SALI fluctuates around a non-zero value for
ordered orbits, while it tends to zero for chaotic
orbits.\cite{Sk01,GRACM} In the case of 2D maps the SALI tends to
zero both for ordered and chaotic orbits, following however
completely different time rates, which again allows us to
separate between these two cases also.\cite{Sk01}

We have recently begun to understand the different behaviors of
the SALI in regions of order and chaos. In the latter case, we
have been able to connect SALI's rapid convergence to zero, to the
influence of the two largest positive Lyapunov exponents of the
motion.\cite{prep} In the present paper we shall study the
behavior of the SALI in the case of ordered orbits.

\section{The behavior of the SALI for ordered motion}

Let us try to understand why the SALI  does not become zero in the
case of ordered motion, by studying in detail the behavior of the
deviation vectors. A suitable way to do this for conservative
systems is to consider a non-trivial integrable Hamiltonian model
whose orbits are bounded and lie on ``nested'' tori, which foliate
all of the available phase space.\cite{LL}

An integrable such Hamiltonian system of 2 degrees of freedom
possesses besides the Hamiltonian  $H$  a second independent
integral  $F$, in involution with~$H$:
\begin{equation}
\{ H,F \}=0, \label{eq:HF}
\end{equation}
where $\{ \cdot , \cdot \} $ denotes the usual Poisson bracket. In
such systems, the motion lies in the intersection of  both
manifolds
\begin{equation}
H=\widetilde{h}, \,\,\, F=\widetilde{f}, \label{eq:man}
\end{equation}
where $\widetilde{h}$, $\widetilde{f}$ are the constant values of
the two integrals. Thus, the orbits in the 4--dimensional phase
space move instantaneously on a 2--dimensional ``tangent''
subspace, which is `perpendicular' to the vectors
\begin{equation}
\overrightarrow{\nabla H}= (H_x,H_y,H_{p_x}, H_{p_y}), \,\,\,
\overrightarrow{\nabla F}= (F_x,F_y,F_{p_x}, F_{p_y}),
\label{eq:grads}
\end{equation}
$x$, $y$ being the generalized coordinates of the system and
$p_x$, $p_y$ their conjugate momenta, while  subscripts denote
partial derivatives (e.\ g.\ $H_x \equiv \frac{\partial
H}{\partial x}$). In fact, the motion may be thought of as
governed by either one of the Hamiltonian vector fields
\begin{equation}
\overrightarrow{f_H}= (H_{p_x}, H_{p_y},-H_x,-H_y), \,\,\,
\overrightarrow{f_F}= (F_{p_x}, F_{p_y},-F_x,-F_y).
\label{eq:flow}
\end{equation}
The vectors $\overrightarrow{\nabla H}$, $\overrightarrow{\nabla
F}$ (and hence also $\overrightarrow{f_H}$,
$\overrightarrow{f_F}$) are linearly independent due to the
functional independence of the two integrals at almost all points
in phase space. So the corresponding unit vectors
\begin{equation}
\widehat{f_H} =
\frac{\overrightarrow{f_H}}{\|\overrightarrow{f_H}\|} \bot
\widehat{\nabla H}, \,\,\, \widehat{f_F} =
\frac{\overrightarrow{f_F}}{\|\overrightarrow{f_F}\|} \bot
\widehat{\nabla F}, \,\,\, \mbox{ with } \,\,\, \widehat{\nabla H}
= \frac{\overrightarrow{\nabla H}}{\|\overrightarrow{\nabla H}\|},
\,\,\, \widehat{\nabla F} = \frac{\overrightarrow{\nabla
F}}{\|\overrightarrow{\nabla F}\|} \label{eq:base}
\end{equation}
can be used as a basis for the 4--dimensional  space where the
deviation vectors evolve. This basis is in general not orthogonal
as
\begin{equation}
\langle \widehat{\nabla H}, \widehat{\nabla F} \rangle = \langle
\widehat{f_H}, \widehat{f_F} \rangle = \frac{H_x F_y + H_y F_y
+H_{p_x} F_{p_x} + H_{p_y} F_{p_y}} {\|\overrightarrow{\nabla H}\|
\,\|\overrightarrow{\nabla F}\|} \label{eq:dots1}
\end{equation}
is not necessary zero. We note that $\|\overrightarrow{\nabla H}\|
=\|\overrightarrow{f_H}\|$, $\|\overrightarrow{\nabla F}\|
=\|\overrightarrow{f_F}\|$ and $\langle \cdot , \cdot \rangle$
denotes the usual inner product. Note also that from definitions
(\ref{eq:grads}) and (\ref{eq:flow}) we get $\langle
\widehat{\nabla H}, \widehat{f_H} \rangle = \langle
\widehat{\nabla F}, \widehat{f_F} \rangle = 0$, while
(\ref{eq:HF}) yields $\langle \widehat{\nabla H}, \widehat{f_F}
\rangle = \langle \widehat{\nabla F}, \widehat{f_H} \rangle = 0.$

So, using vectors (\ref{eq:base}) as a basis for studying the
evolution of a deviation vector $\overrightarrow{v_1}$, we can
write it as
\begin{equation}
\overrightarrow{v_1} =
a_1 \widehat{f_H} + a_2 \widehat{f_F} +
a_3 \widehat{\nabla H} + a_4  \widehat{\nabla F}
\label{eq:vector}
\end{equation}
with $a_1, \, a_2, \, a_3, \, a_4 \in \mathbb{R}$. The values of
the coefficients $a_i$, $i=1,2,3,4$, at different times, give us a
clear picture for the evolution of $\overrightarrow{v_1}$. In the
case of the 2D standard map for example, where ordered orbits lie
on an invariant curve (1D torus), it has been shown both
numerically and analytically\cite{Voz}  that any deviation vector
(considered as a linear combination of the vectors
$\widehat{f_H}$, $\widehat{\nabla H}$ using our notation),
eventually becomes tangent to the invariant curve, tending to the
tangential direction as $n^{-1}$, with $n$ being the number of
iterations.

Similarly, in the case of an integrable 2D Hamiltonian the
deviation vector $\overrightarrow{v_1}$ tends to fall on the
``tangent space'' of the torus, spanned at each point by
$\widehat{f_H}$, $\widehat{f_F} $, meaning that in
Eq.~(\ref{eq:vector}) $a_3 \rightarrow 0$, $a_4 \rightarrow 0$,
while the $a_1$, $a_2$ are, in general, different from zero. This
is analogous to what has been found for the 2D standard map
 in Ref.~\citen{Voz}. As a model for studying this behavior
let us consider the 2D Van der Waals Hamiltonian \cite{kn:1}
\begin{equation}
H(x,y,p_{x},p_y)=
\frac{1}{2}(p_{x}^{2}+p_y^{2})-E(x^2+y^2)+A(x^{6}+y^6)+
B(x^{4} y^2+x^{2} y^{4}),
\label{eq:Ham1}
\end{equation}
where $E$, $A$, $B$ are real parameters. For $B=3A$ and
$E\in\mathbb{R}$  the Hamiltonian (\ref{eq:Ham1}) is completely
integrable and the second integral of motion is given
by\cite{kn:1}
\begin{equation}
F(x,y,p_{x},p_y)=(x p_y-y p_{x})^{2}.
\label{eq:Fint}
\end{equation}
In our calculations we consider the integrable case for $A=0.25$,
$B=3A=0.75$ and $E=-10^{-8}$.

\begin{figure}
\centerline{\includegraphics[width=15 cm,height=7.5 cm]
{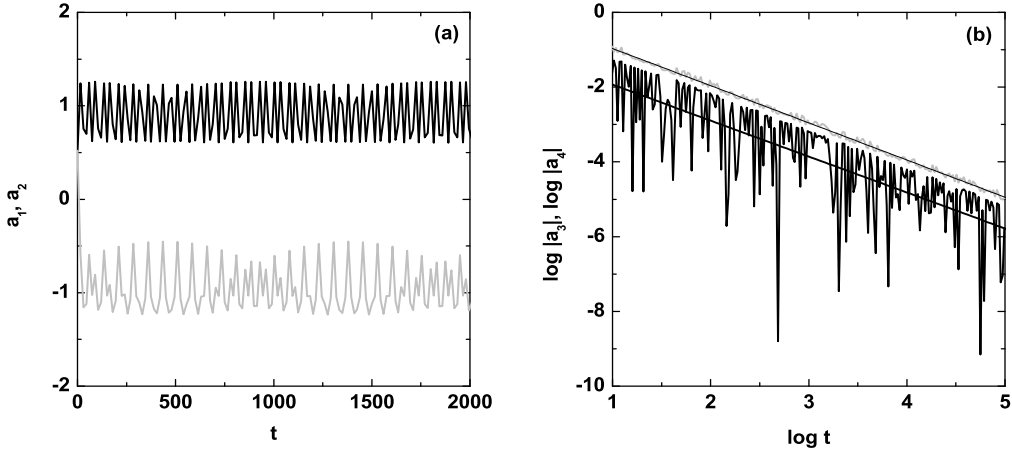}} \caption{ The time evolution of the coefficients
$a_1$, $a_2$, $a_3$, $a_4$, of the initial deviation vector with
$a_1=1$, $a_2=1$, $a_3=0$, $a_4=1$ (Eq.~(\ref{eq:vector})). (a)
$a_1$ (black line), $a_2$ (gray line). (b) $|a_3|$ (black line),
$|a_4|$ (gray line) in log-log scale. The orbit of the Hamiltonian
(\ref{eq:Ham1}) used, has initial condition $x=-0.6$, $y=0$,
$p_x=0$, $p_y=1.99416$. In (b) we also plot the curves
$|a_3|=0.107 \cdot t^{-0.962}$, $|a_4|=1.067 \cdot t^{-0.995}$
that fit the data.} \label{fig:1}
\end{figure}

\begin{figure}
\centerline{\includegraphics[width=15 cm,height=7.5 cm]
{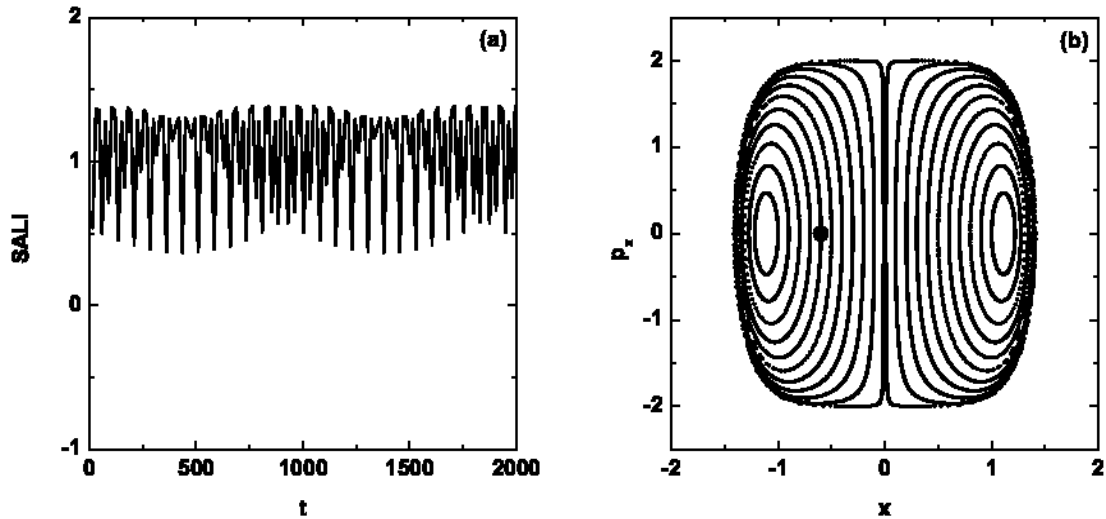}} \caption{The time evolution of the SALI (a) for the
ordered orbit marked by a black point in the PSS $y=0$ of the
system (b).} \label{fig:2}
\end{figure}

For different initial deviation vectors, we compute the time
evolution of the coefficients $a_1$, $a_2$, $a_3$, $a_4$ of
Eq.~(\ref{eq:vector}). We find that in all cases $a_1$, $a_2$
remain different from zero, while $a_3$, $a_4$ tend to zero. A
particular example is given in Fig.~\ref{fig:1}. By fitting the
data of Fig.~\ref{fig:1}b we see that $|a_3|$, $|a_4|$ $\propto
t^{-1}$. From Fig.~\ref{fig:1} we conclude  that any vector will
eventually fall on the ``tangent space'' of the torus on which the
orbit evolves. This ``tangent space'' is produced by vectors
$\widehat{f_H}$,  $\widehat{f_F}$, and so any deviation vector
will eventually become a linear combination of these two vectors
only. As there is no particular reason for  two different initial
deviation vectors to end up with the same values of $a_1$, $a_2$,
the SALI (\ref{eq:SALI}) will in general oscillate around a value
different from zero. In other words the two vectors become tangent
to the torus and fluctuate quasiperiodically about two different
directions. This  becomes evident in Fig.~\ref{fig:2}a where we
plot the time evolution of the SALI for an orbit with initial
conditions $x=-0.6$, $y=0$, $p_x=0$, $p_y=1.99416$ marked by a
black point in the Poincar\'{e} Surface of Section (PSS) of the
system seen in Fig.~\ref{fig:2}b. The initial deviation vectors
used are $\overrightarrow{v_1}= \widehat{f_H} + \widehat{f_F}  +
\widehat{\nabla F}$ (the time evolution of which is given in
Fig.~\ref{fig:1}) and $\overrightarrow{v_2}=\widehat{f_H} +
\widehat{f_F} + \widehat{\nabla H} $.

\section{Conclusions}
In this paper we have analyzed the behavior of the SALI in regions
of ordered motion, by studying the evolution of deviation vectors
in the case of an integrable 2D Hamiltonian system. Using a
suitable basis of 4 vectors  (\ref{eq:base}) we have shown that
any pair of arbitrary deviation vectors tends to the tangential
space of the torus, following a $t^{-1}$ time evolution and having
in general 2 different directions. This explains why for ordered
orbits the SALI oscillates quasiperiodically about values that are
different from zero. The same result is observed to hold in the
case of ordered motion in a stability region of a non--integrable
system,\cite{GRACM} where the presence of  ``islands'' implies the
existence of an additional approximate integral $F$, independent
of the Hamiltonian. For Hamiltonian systems of more than 2 degrees
of freedom we expect similar results. The only difference is that
the ``tangent space'' is of higher dimension generated by the
vectors $\widehat{f_H},
\,\,\widehat{f_{F_1}},\,\,\widehat{f_{F_2}}, \,\ldots$, with
$F_1$, $F_2, \, \ldots$, being the additional (approximate or not)
integrals of the motion.

\hyphenation{Ka-ra-the-o-do-ry}
\section*{Acknowledgements}
Ch.\ Skokos was  supported by the `Karatheodory' post--doctoral
fellowship No 2794 of the University of Patras and by the Research
Committee of the Academy of Athens (program No 200/532). Ch.\
Antonopoulos was supported by the `Karatheodory' graduate student
fellowship No 2464 of the University of Patras.

\end{document}